\documentclass[twocolumn,prl,floatfix,superscriptaddress]{revtex4}
\usepackage{graphicx, subfigure, amsfonts,amssymb,amsmath, array, gensymb,fontenc,color, lipsum}

\begin{document}

\title{ {Magnetic exchange interactions in monolayer CrI$_3$ \\ from many-body wavefunction calculations}}

\author{Michele Pizzochero}
\email{michele.pizzochero@epfl.ch}
\author{Ravi Yadav}
\author{Oleg V.\ Yazyev}
 
\affiliation{Institute of Physics, Ecole Polytechnique F\'ed\'erale de Lausanne (EPFL), CH-1015 Lausanne, Switzerland}
\affiliation{National Centre for Computational Design and Discovery of Novel Materials (MARVEL), Ecole Polytechnique F\'ed\'erale de Lausanne (EPFL), CH-1015 Lausanne, Switzerland}

\date{\today}

\begin{abstract}
The marked interplay between the crystalline, electronic, and magnetic structure of atomically thin magnets has been regarded as the key feature for designing next-generation magneto-optoelectronic devices. In this respect, a detailed understanding of the microscopic interactions  underlying the magnetic responses of these crystals is of primary importance. Here, we combine model Hamiltonians with multi-reference configuration interaction wavefunctions to accurately determine the strength of the spin couplings in the prototypical single-layer magnet CrI$_3$. Our calculations identify the (ferromagnetic) Heisenberg exchange interaction $J = -1.44$ meV as the dominant term, being the inter-site magnetic anisotropies substantially  {weaker}. We also find that single-layer CrI$_3$ features an out-of-plane easy axis ensuing from a single-ion anisotropy  $A = -0.10$ meV, and predict $g$-tensor in-plane components $g_{xx} = g_{yy} = 1.90$ and out-of-plane component $g_{zz} = 1.92$. In addition, we assess the performance of a dozen widely used density functionals against our accurate correlated wavefunctions  {calculations} and available experimental data, thereby establishing reference results for future first-principles investigations.  Overall, our findings offer a firm theoretical ground to experimental observations.
\end{abstract}

\maketitle

\section{Introduction} 

The first isolation of graphene back in 2004 represented a paradigm shift in condensed matter physics \cite{Novoselov04}.
This discovery sparked a wealth of unexplored research directions, most notably the search for novel two-dimensional
crystals \cite{Novoselov05, Nicolosi13, Mounet18}, together with their controlled stacking for the assembling of van der Waals heterostructures in a layer-by-layer fashion \cite{Novoselov16, Geim13, Chen2018}.
Since then, the library of atomically thin materials  {is rapidly expanding}, first starting from those parent 3D crystals 
which can be peeled off to the 2D limit (\emph{e.g.}\ transition metal dichalcogenides \cite{Manzeli17} or hexagonal boron nitride \cite{Li16}) and 
subsequently evolving into the synthesis of artificial monolayers (\emph{e.g.}\ silicene \cite{Vogt12, Tao2015, Piz16} or its heavier group IV analogs \cite{Zhu2015, Pizzochero19}), eventually covering 
a broad range of properties, including metals \cite{CastroNeto09, Ciarrocchi2018}, semi- and super-conductors \cite{Mak10, Yazyev15, MWChen17, CastroNeto01, Cao18a}, as well as trivial or topological insulators \cite{Qian14, Tang17, Pedramrazi19}.

Of particular interest in this context is the emergence of correlated electronic phases in ultrathin nanostructures, 
being the observation of magnetically ordered phases in some of these systems arguably the most striking of such phenomena \cite{CastroNeto01, Cao18a, Cao18b}.  
Indeed, while magnetism in lattices of reduced dimensionalities has hitherto been ascribed to native or engineered impurities \cite{Yazyev07, Nair2012, Pizzochero17, Avsar19}, 
the isolation and characterization of two-dimensional CrI$_3$ (along with other ultrathin magnets, \emph{e.g.}\ Fe$_3$GeTe$_2$, Cr$_2$Ge$_2$Te$_6$, VSe$_2$, FePSe$_3$, to mention but a few)  have unambiguously demonstrated the realization of a long-range, intrinsic magnetism in atomically thin crystals \cite{Huang2017, Burch2018, Gibertini19, Gong19, Mak2019, Torelli2019}. This achievement is made possible as a consequence of the pivotal role played by the magnetocrystalline anisotropy, which preserves the magnetic order down to the monolayer limit, as explained by the Mermin-Wagner theorem \cite{Mermin66}.  Experimentally, ultrathin films of CrI$_3$ have been obtained upon exfoliation of their bulk counterpart, an insulating layered crystal exhibiting magnetic order up to 61 K \cite{Huang2017}. Interestingly,  {a competition between coexisting intra-layer ferromagnetic and inter-layer antiferromagnetic exchange interactions in few-layer samples has been unraveled, giving rise to a thickness-dependent magnetic response \cite{Huang2017}. In addition, switching between these two exchange interactions} can widely be engineered in atomically thin samples, \emph{e.g.}\ through applied external pressure \cite{Song19, Tingxin19}, electrostatic-gate control \cite{Huang2018, Jiang2018}, or lattice deformations \cite{Thiel19}. Altogether, this pronounced interplay between crystalline, magnetic, and electronic structures paves the way towards prospective magneto-optoelectronic devices based on thin films of CrI$_3$ \cite{Wang2018, Song1214, Klein1218, McGuire15, Mak2019}. In this vein, establishing the microscopic spin physics governing this system is of paramount importance. 

On the computational side, the accurate description of magnetic interactions in two-dimensional CrI$_3$ comes as a challenging task, mainly due to the inherent inadequacy of semilocal density functionals in properly capturing electron-electron interactions. 
Several approximations have been devised for describing localization effects (\emph{e.g.} Hubbard-corrected or hybrid density functionals), but they invariably rely on adjustable parameters, which in turn are system- and property-dependent and cannot be determined following universal protocols. These limitations can effectively be overcome by relying on many-body wavefunctions, which, though computationally demanding, enable to recover a substantial amount of the correlation energy. In addition, we stress that the nature and strength of numerous magnetic interactions occurring in monolayers of CrI$_3$ yet remain largely unknown, and their determination is essential in order to provide a detailed comprehension of the intriguing spin physics hosted by this crystal.  Here,  {we report on the first \emph{ab initio} quantum chemistry} investigation of magnetic interactions in two-dimensional CrI$_3$ by carrying out  {multi-reference configuration interaction calculations.} We further exploit such benchmark results to assess the performances of several density-functional approximations. Overall, our work portrays an unprecedentedly accurate picture of the spin interactions in monolayer CrI$_3$, which is instrumental in understanding its magnetic properties.

\section{Crystal and electronic structure} 

\begin{figure}[]
  \centering
  \includegraphics[width=0.95\columnwidth]{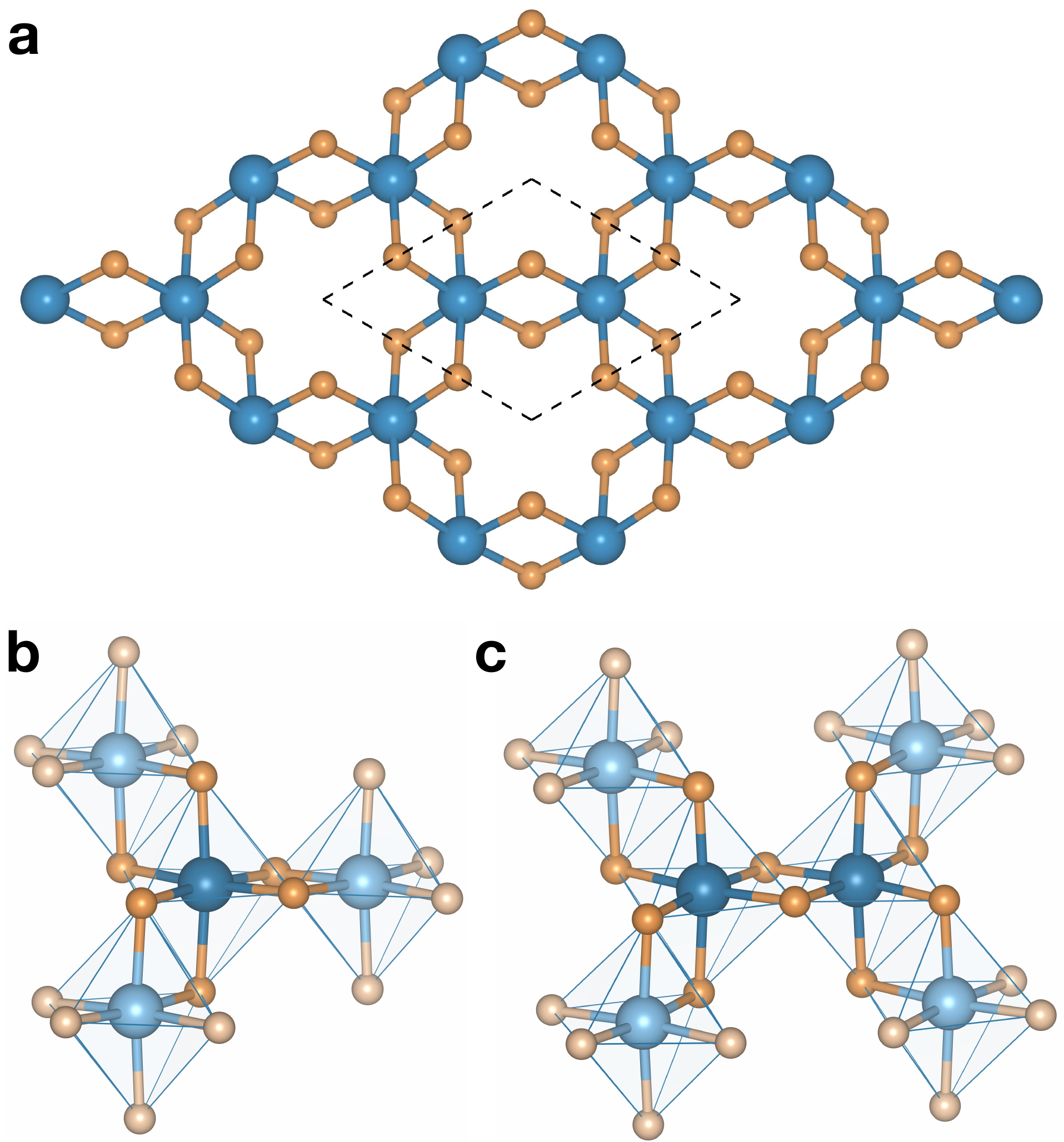}
  \caption{ {(a) Crystal structure of single-layer CrI$_3$. Blue and orange balls represent chromium and iodine atoms, respectively. The unit cell is indicated with black dashed lines. (b) One-site and (c) two-site finite-size models adopted for the quantum chemistry calculations. The atoms shown in darker (lighter) colors are treated at the correlated (Hartree-Fock) level. The fragments are further embedded in a periodic array of point charges (not shown). } \label{Fig1}}
\end{figure}

We start by briefly reviewing the intertwinement  between the crystalline and  electronic structure of  CrI$_3$. Down to the monolayer limit, CrI$_3$ consists of a honeycomb plane of Cr atoms sandwiched between two planes of I atoms, as shown in Fig.\ \ref{Fig1}. Each Cr atom exhibits a six-fold coordination, which gives rise to edge-sharing octahedra. According to a purely ionic argument, the Cr atoms present a formal oxidation state of +3 and a resulting valence electron configuration $3d^3 4s^0$. As a consequence of the crystal field associated with the octahedral environment, a splitting of the $d$ orbitals into three triply occupied $t_{2g}$ states and two higher-energy empty $e_g$ states occurs, the extent of which has not been ascertained yet and will be given in the following. However, on the basis of the Hund's rule, such an electron occupation yield $S = 3/2$ \cite{Lado17}. This has been experimentally confirmed in single- and multilayer CrI$_3$, where a magnetization saturation of 3 $\mu_B$ per Cr$^{3+}$ ion has been observed \cite{McGuire15, Huang2017}.

The scenario mentioned above suggests that Cr$^{3+}$ ion act as magnetic centers, which interact \emph{via} non-magnetic iodine ligands through the so-called super-exchange coupling theoretically proposed by P.W.\ Anderson \cite{Anderson50}. On the experimental side, the hysteretic features emerging in magneto-optical Kerr effect (MOKE) measurements of single-layer CrI$_3$ are the hallmarks of a ferromagnetic spin order \cite{Huang2017},  which, given the Cr-I-Cr bond angle of $\sim$90\textsuperscript{o}, is consistent with the Goodenough-Kanamori rule \cite{Goodenough58, Kanamori59}. Furthermore, such experimental investigations revealed that single-layer CrI$_3$ displays an out-of-plane easy axis and a critical temperature  of 45 K, only slightly lower than its three-dimensional counterpart \cite{Huang2017}. In the following, we accurately quantify the magnetic interactions underlying these effects.

\section{Spin Hamiltonian from many-body wavefunctions} 

 We consider a generalized, bilinear model Hamiltonian, which captures intra-  as well as inter-site between the $i$-th and $j$-th nearest-neighbor centers  {with} spins $\vec{S}_i$ and $\vec{S}_j$, respectively. Such a Hamiltonian reads as
\begin{equation}
\begin{split}
 \mathcal{H} = J  \sum_{i,j} \vec{S_i} \cdot \vec{S_j} + \sum_{i,j} \vec{S_i} \cdot \bar{\bar{\Gamma}} \cdot \vec{S_j}  +  \sum_{i}AS_i^2 +  \\ +  \sum_i \mu_B \vec{B} \cdot \bar{\bar{g}} \cdot \vec{S_i} + \sum_{i,j}  \vec{D} \cdot \vec{S}_i \times \vec{S}_j 
\end{split}
\label{Ham}
\end{equation}
with $J$ corresponding to the isotropic Heisenberg exchange, $\bar{\bar{ \Gamma}}$ the symmetric anisotropic tensor,  $A$ the single-ion anisotropy parameters,  $\bar{\bar{g}}$ the $g$-tensor in the Zeeman term that accounts for the interaction of the lattice with the external magnetic field $\vec{B}$, and $\vec{D}$ the antisymmetric Dzyaloshinsky-Moriya interaction parameter. This latter term vanishes due to the $C_{2h}$ point group of the CrI$_3$ lattice and the ensuing inversion symmetry. Assuming a local Kitaev frame according to which the $z$ axis is perpendicular to the Cr$_2$I$_2 $ plaquette for each Cr-Cr bond \cite{katukuri14,yadav16,yadav18}, $\bar{\bar{\Gamma}}$ takes the form
 \begin{equation}
 \bar{\bar{\Gamma}} = 
\begin{pmatrix} 
0 & \Gamma_{xy} & -\Gamma_{yz} \\ 
\Gamma_{xy}  & 0 & \Gamma_{yz}  \\ 
-\Gamma_{yz} & \Gamma_{yz} & K
\end{pmatrix}
 \end{equation}
 with $K$ being the Kitaev  {parameter}.
 
In order to determine the magnetic interactions contained in Eq.\ (\ref{Ham}), we perform many-body wavefunction  calculations on carefully chosen embedded model systems. The models consist of a central unit containing either one [Fig.\ \ref{Fig1}(b)] or two [Fig.\ \ref{Fig1}(c)] edge-sharing octahedra treated at the correlated level, surrounded by the nearest neighbor octahedra [see Fig. \ref{Fig1}(b)], the orbitals of which are frozen at the Hartree-Fock level. To ensure charge neutrality and mimic the periodic environment to which the finite-size model is subjected in the extended system, the fragment is embedded in an array of point charges fitted to the reproduce the Madelung ionic potential of the crystalline lattice. We rely on the model shown in Fig.\ \ref{Fig1}(b) and Fig.\ \ref{Fig1}(c)  to determine intra- and inter-site magnetic interactions, respectively. Electron correlation effects in the central unit are described at both the 
{  complete-active-space self-consistent-field (CASSCF) as well as multi-reference configuration interaction (MRCI) levels of theory \cite{Helgaker2000}, including spin-orbit interactions. As a first step, multi-configuration reference
wavefunctions are constructed for an active space spanned by six electrons in six $t_{2g}$ levels (that is, 3 at each Cr site). The orbitals are optimized for an average of low-lying septet, quintet, triplet, and singlet states. Next, MRCI calculations are performed including single and double excitations involving the $d$ ($t_{2g}$) valence shells of Cr$^{3+}$ ion and the $p$ valence shells of the  bridging I ligands ({see Appendix A}) \cite{Helgaker2000}}. We anticipate that, while inessential for the intra-site magnetic interactions, an MRCI treatment of the electron correlation is crucial to accurately determine the inter-site magnetic parameters due to the important role of the super-exchange coupling, being this latter effect neglected in the CASSCF wavefunctions. Finally, we quantify the nature and magnitude of the magnetic parameters in CrI$_3$ by mapping the resulting \emph{ab initio} Hamiltonian onto the model Hamiltonian of Eq.\ (\ref{Ham})  through the well-established procedure detailed in Ref. [\onlinecite{Bogdanov13}].

\begin{table}[]
\caption{\label{tab:table1}%
 {Relative energies of the Cr$^{3+}$ 3$d^3$ multiplet structure obtained with the many-body wavefunction CASSCF and MRCI methods on finite-size model systems. Energies are given in eV.}
}
\begin{ruledtabular}
\begin{tabular}{ c c c }
\textrm{ 3$d^3$ splitting}&
\textrm{CASSCF} &
\textrm{MRCI}\\
\colrule
$^4A_2 (t_{2g}^3 e_{g}^0)$ &  0.00  				&  0.00  \\
$^4T_2 (t_{2g}^2 e_{g}^1)$ &  1.48; 1.49; 1.49      	&  1.62; 1.67; 1.67 \\
$^2E (t_{2g}^1 e_{g}^2)$ &  2.34; 2.34  	 		& 2.22; 2.22 \\
$^4T_2 (t_{2g}^2 e_{g}^1)$ & 2.35; 2.42; 2.42       	& 2.50; 2.58; 2.58 \\
$^2T_2 (t_{2g}^1 e_{g}^2)$ & 2.45; 2.46; 2.46       	& 2.33; 2.34; 2.34 
\label{Tab1}
\end{tabular}
\end{ruledtabular}
\end{table}

Before moving to spin interactions, we briefly elucidate the electronic structure of the magnetic centers in single-layer CrI$_3$.  The relative energies of the multiplet structure of the 3$d^3$ orbitals of the Cr$^{3+}$ ion are given in Table \ref{Tab1}. As discussed above, according to the crystal-field theory, one should expect a singlet ground state for the Cr ion residing in an octahedral environment. This is confirmed by our correlated wavefunction calculations, which found $^4A_2(t_{2g}^3 e_{g}^0)$  to be the lowest-energy configuration. The higher-energy terms $^4T_2(t_{2g}^2 e_{g}^1)$ and $^2E (t_{2g}^1 e_{g}^2)$  can interact with the $^4A_2(t_{2g}^3 e_{g}^0)$  ground state in the presence of spin-orbit coupling, {  inducing a zero-field splitting of the spin quartet into $m_s= \pm 1/2$ and $m_s=\pm 3/2$ Kramers doublets, respectively}. Our CASSCF results obtained including all 3$d$ orbitals at the Cr site in the active space indicate that the $^4T_2(t_{2g}^2e_g^1)$ configuration lies $1.48 - 1.49$ eV higher in energy than the $^4A_2$ one.  {This quantity slightly increases to $1.62 - 1.67$ eV at the MRCI level. We notice that the spin-orbit interactions affect the splittings given in Table \ref{Tab1} by only $\sim$0.1 meV.}

The inter- and intra-site magnetic parameters of Eq.\ (\ref{Ham}) obtained at both the CASSCF and MRCI levels are listed in Table \ref{Tab2}. As far as the inter-site interactions are concerned, we find a dominant isotropic Heisenberg exchange $J = -1.44$ meV, signaling a ferromagnetic ground state of single-layer CrI$_3$, in line with experimental observations \cite{Huang2017} and earlier theoretical results.  {Indeed, this value lies in between those previously reported, which range from $-1.10$ meV to $-1.63$ meV \cite{Lado17, Xu2018, Torelli18}.} Such an isotropic Heisenberg coupling largely exceeds the anisotropic exchange interactions, being the Kitaev parameter contributing only up to $-0.08$ meV and off-diagonal anisotropic terms smaller than $J$ by several orders of magnitude.   {This finding is at odds with the first-principles results of Ref.\ \cite{Xu2018}, in which a Kitaev interaction of magnitude comparable to $J$ was reported.}   {It is worth noticing that the CASSCF method largely underestimates the inter-site exchange couplings, as compared to the MRCI method. This is especially true for the isotropic Heisenberg exchange, which at the CASSCF level is found to be 0.82 meV lower than that obtained at the MRCI level. This difference points towards (and, to some extent, quantifies) the significant role of the super-exchange channels between the Cr$^{3+}$ ions occurring \emph{via} the bridging I ligands in governing the magnetism of single-layer CrI$_3$.}

We next discuss intra-site magnetic interactions.  {  Even in the absence of magnetic field, the interplay between the spin-orbit coupling and the crystal field lifts the degeneracy of the electronic ground state for $S>1/2$. The extent of such a zero-field splitting is quantified by the single-ion anisotropy.} We derive this quantity by following the effective-Hamiltonian methodology presented in Ref. \cite{Maurice09}. In this approach, the mixing of the  $^4A_2$ components with the higher-energy states is treated in a perturbative manner, and the spin-orbit wavefunctions related to the high-spin $t_{2g}^3$ configurations are projected onto the $^4A_2$ $|S,M_s\rangle$ states. We then construct the effective Hamiltonian $\tilde{H}_{\rm eff}=\Sigma_k E_k |\tilde{\psi}_k\rangle \langle \tilde{\psi}_k|$, where $\tilde{\psi}_k$ are the ortho-normalized projections of the low-lying quartet wavefunctions with corresponding eigenvalues $E_k$.
A one-to-one correspondence between $\tilde{H}_{\rm eff}$ and the model Hamiltonian $\tilde{H}_{\rm mod}={S\cdot \bar{\bar{A}} \cdot S}$ leads to the ${ \bar{\bar{A}}}$ tensor, which, upon diagonalization, yields the commonly used axial parameter $A$ \cite{Maurice09}.
We obtain $A = -0.10$ meV, with the magnetic axis lying in the direction normal to the lattice plane. The negative sign of $A$ indicates that the easy axis of magnetization points along the magnetic axis, in accord with experiments \cite{Huang2017}.  {We remark that this value is less than half than that obtained in Ref.\ \cite{Xu2018} by means of density functional theory.}

Finally, we quantify the response of the magnetic Cr sites to the external magnetic field $\vec{B}$ through the determination of the $g$-tensor appearing in the Zeeman term of the spin Hamiltonian in Eq.\ (\ref{Ham}). This quantity can readily be accessed in experiments, \emph{i.e.}\ electron spin resonance measurements  {\cite{epr}}. {  From the quantum chemistry point of view, the multi-configuration wavefunctions are known to provide an accurate description of the spin-orbit multiplets. This allows one to evaluate the matrix elements of the total magnetic moment operator ${\bf \hat{\mu}}=-\mu_B(g_e\hat{{S}}+ \hat{{L}})$ in the basis of the multiplet eigenstates, with $\hat{{L}}$ and $\hat{{S}}$ being the angular momentum and spin operators, whose expectation values are obtained for a given Cr site, and $g_e$ being  the free-electron Land\'e factor. The Zeeman Hamiltonian can be written in terms of total moment $\mu$ as $H_{Z}=-\hat{{\mu}} \cdot \vec{B}$. In order to obtain the $g$-tensor, such $H_{Z}$ is mapped onto the Zeeman model Hamiltonian presented in Eq.\ (\ref{Ham}) through the well-established procedure devised in Refs.\ \cite{Bolvin06,Chibo2}. To this end, we rely on an active space that encompasses all the five $d$ orbitals of Cr (with 3 electrons). We construct the initial wavefunction by averaging over the seven quartets and five doublets listed in Table \ref{Tab1}.}
We find an anisotropic $g$-tensor, with  $g_{xx}=g_{yy}=1.90$ and $g_{zz}=1.92$, where the $x=y$ ($z$) axis lies in the (perpendicular to) the CrI$_3$ lattice plane.

\begin{table}[]
\caption{\label{tab:table1}%
 {Magnetic exchange coupling parameters} in single-layer CrI$_3$: isotropic Heisenberg magnetic exchange ($J$), symmetric anisotropic ($\Gamma_{xy}$, $\Gamma_{yz}$, and $\Gamma_{zx}$), and  Kitaev ($K$) interactions along with single-ion anisotropy ($A$) and  {the components of the $g$-tensor ($g_{xx}$, $g_{yy}$, $g_{zz}$)} calculated by means of CASSCF and MRCI methods.}
\begin{ruledtabular}
\begin{tabular}{ l c c }
\textrm{Magnetic exchange  {coupling}}&
\textrm{CASSCF} &
\textrm{MRCI}\\
\colrule
$J$ (meV)			  & --0.62  	&  --1.44  \\
$K$ (meV)	 			  &  --0.01      	&  --0.08 \\
$\Gamma_{xy}$ (meV)	 & --1.0 $\times$ 10$^{-3}$  	& --2.3 $\times$ 10$^{-3}$\\
$\Gamma_{yz} = -\Gamma_{zx}$ (meV)	  & --2.1 $\times$ 10$^{-4}$ & 	--1.2 $\times$ 10$^{-3}$ \\
$A$ (meV)	 & 	--0.10 &   \\
$g_{xx} = g_{yy}$ &  1.90	& \\
$g_{zz}$ &  1.92	& \\
\end{tabular}
\end{ruledtabular}
\label{Tab2}
\end{table}

\section{Magnetic interactions from density functional theory} 

With the accurate many-body wavefunction results at hand, we are in a position to assess the performance of widespread-used exchange and correlation functionals in describing \emph{selected} magnetic interactions occurring in single-layer CrI$_3$. We address by means of first-principles calculations in a periodic setting the isotropic Heisenberg exchange $J$, the magnetic anisotropy $E\textsubscript{MAE}$, and provide an estimate of the Curie temperature ($T_C$), for which the experimental value (45 K) is available. Specifically, we climb the ladder of Density Functional Theory (DFT) by considering the local density approximation  {using the Ceperley and Alder (CA) parametrization} \cite{LDA}, several flavors of the generalized gradient approximation (PBE \cite{PBE}, PBEsol \cite{PBEsol}, PW91 \cite{PW91}, revPBE \cite{revPBE}) and its Hubbard-corrected extension DFT+$U$ \cite{DFTU} (with 1.0 eV $\leq U \leq$ 3.0 eV), some representative examples of meta generalized gradient approximations (SCAN \cite{SCAN}, TPSS, and RTPSS \cite{RTPSS}), and hybrid Fock-exchange/density-functionals, both in their plain (PBE0 \cite{PBE0}) and range-separated (HSE03 \cite{HSE03} and HSE06 \cite{HSE06}) formalisms.  {Further details of our DFT calculations are provided in Appendix B.}

The main finding of our MRCI investigation is that, among the inter-site interactions listed in Table \ref{Tab2}, the dominant one is the Heisenberg exchange coupling. This indicates that single-layer CrI$_3$ can effectively be described as an isotropic Heisenberg magnet. Hence, we map the first-principles results onto the isotropic Heisenberg spin Hamiltonian $ \mathcal{H} = J  \sum_{i,j} \vec{S_i} \cdot \vec{S_j}$ by determining $J$ according to the usual expression \cite{Whangbo03}
 \begin{equation}
 J = \frac{E\textsubscript{FM} - E\textsubscript{AFM}}{2N\textsubscript{nn}S^2}
  \end{equation}
where $E\textsubscript{FM}$ ($E\textsubscript{AFM}$) is the total energy of the out-of-plane ferromagnetic (antiferromagnetic) phase and $N\textsubscript{nn}$ is the number of the nearest neighbors surrounding the magnetic site.  In order to conduct a meaningful comparison between the many-body wavefunction and the density-functional results, we re-map the \emph{ab initio} Hamiltonian onto the pure Heisenberg Hamiltonian. We obtain a benchmark value of $J = -1.48$ meV at the MRCI level, only slightly larger than that obtained on the basis of Eq.\ (\ref{Ham}). 

\begin{figure}[]
    \includegraphics[width=1\columnwidth]{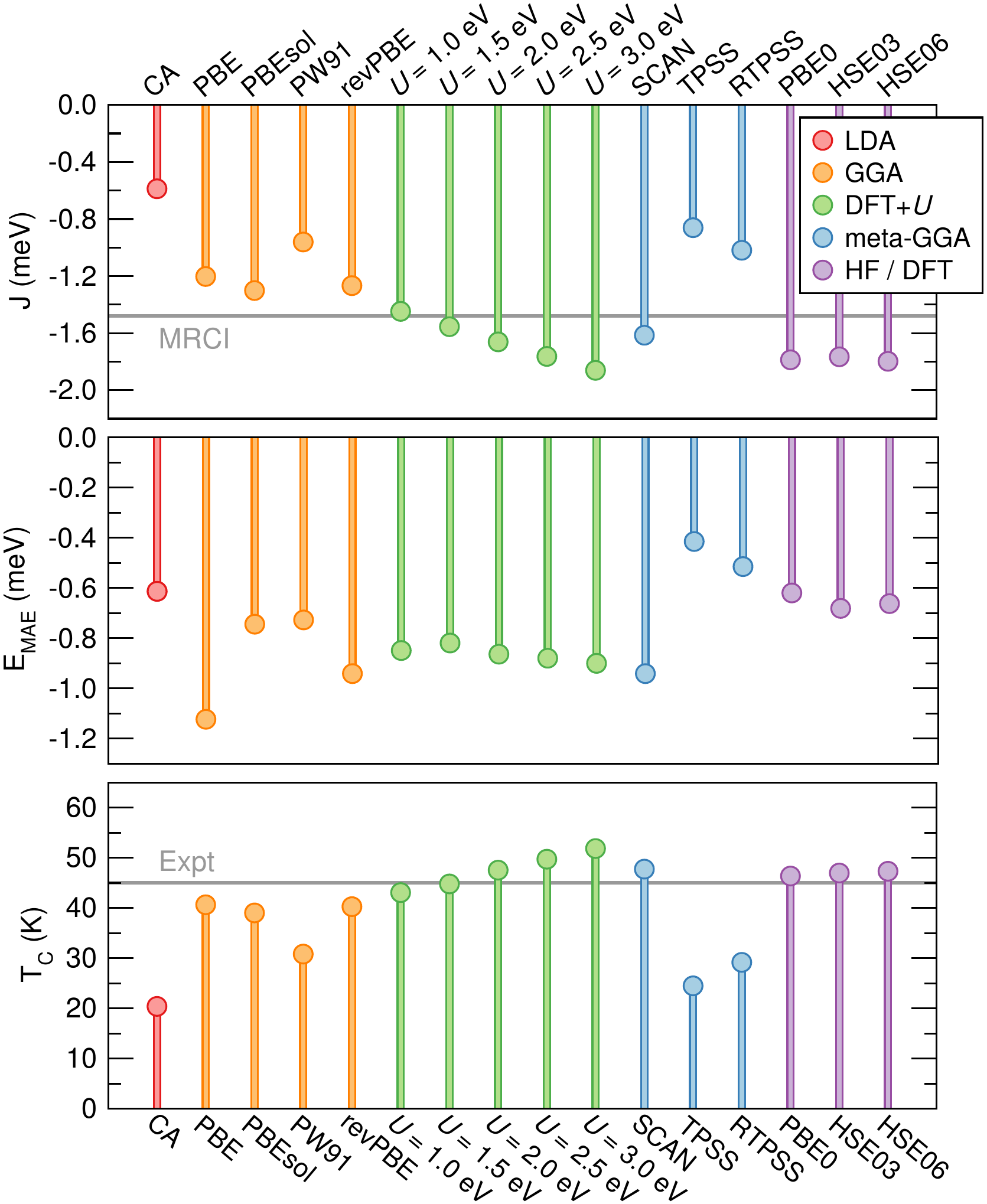}
  \caption{Heisenberg exchange coupling ($J$), magnetocrystalline anisotropy energy ($E\textsubscript{MAE}$), and Curie temperature ($T_C$) of single-layer CrI$_3$ as determined with several classes of density functionals on a periodic model system. Also shown as horizontal black lines are the value of $J$ obtained upon mapping the MRCI Hamiltonian onto a pure Heisenberg Hamiltonian as well as the experimental Curie temperature.}  
  \label{Fig2}
\end{figure}

The upper panel of Fig.\ \ref{Fig2} compares the values of $J$ calculated with several density-functional approximations with our MRCI benchmark result. Notwithstanding their sharp differences in treating electron-electron interactions, all considered functionals provide a qualitative agreement with experiments, yielding a ferromagnetic ground state and an accompanying magnetic moment $\mu$ =  3 $\mu_B$ per Cr$^{3+}$ ion \cite{Huang2017, McGuire15}. However, the magnitude of $J$  largely depends on the adopted approximation. For instance, both CA and (R)TPSS functionals severely underestimate the value of $J$ when compared with the MRCI result, despite their marked dissimilarities in describing exchange and correlation effects. Though to a lesser extent, an underestimation is observed when adopting the generalized gradient approximations as well. Among the meta-GGA functionals considered here, only SCAN is seen to lead to a satisfactory agreement with the benchmark value. Surprisingly, all hybrid functionals  {overestimate} $J$ by  $\sim$20$\%$. The best agreement between DFT and MRCI results is achieved with the PBEsol, SCAN or DFT$+U$  functionals -- the latter matching the MRCI value upon the introduction of moderate values of Coulomb on-site repulsion ($1.0 - 1.5$ eV)  -- as they deviate from the benchmark value by less than 0.2 meV.

Next, we determine the magnetocrystalline anisotropy energy ($E\textsubscript{MAE}$) per formula unit as
\begin{equation}
E\textsubscript{MAE} = E(\mu_\perp) - E(\mu_\parallel)
\end{equation}
with $E(\mu_\perp)$  and  $E(\mu_\parallel)$ being the total energy of single-layer CrI$_3$, with the magnetic moment $\mu$ pointing towards the out-of-plane and in-plane direction, respectively. Our results are overviewed in the middle panel of Fig.\ \ref{Fig2}. MOKE investigations revealed that single-layer CrI$_3$ features an out-of-plane axis, albeit the magnitude of $E\textsubscript{MAE}$ remains undetermined to date. Even though the results of our calculations appear to be spread over a quite broad interval depending on adopted the density-functional, for the ferromagnetic phase we obtain negative magnetocrystalline anisotropy energies irrespectively of the approximation considered. This finding lends support to MOKE observations \cite{Huang2017}, and is further consistent with our many-body wavefunction results concerning the single-ion anisotropy $A$ and $g$-tensor (see Table 2). We then examine the spin orientation of the  {higher-energy} antiferromagnetic phase. As compared to the stable ferromagnetic ordering, this configuration is found to exhibit an in-plane axis and a substantially lower magnetocrystalline anisotropy energy, being $E\textsubscript{MAE} = 0.39, 0.90$, and $0.70$ meV at the PBE, SCAN and HSE06 levels, respectively.

Finally, we address the Curie temperature of CrI$_3$ from first principles and compare the results of our calculations with the experimental value of 45 K. To this end, we rely on the formalism developed in Ref.\ \cite{Torelli18}, in which an analytic expression of $T_C$ in the two-dimensional limit is derived on the basis of a fit to Monte Carlo results achieved on model lattices. In brief, this expression reads as
 \begin{equation}
 T_C = T_I f \left( \frac{\Delta}{J(2S -1)} \right)
 \label{Cur}
 \end{equation}
where  $T_I$ is the critical temperature for the corresponding Ising model  $T_I= S^2 J \tilde{T}_C / k_B$ ($\tilde{T}_C$ = 1.52 in the case of honeycomb lattices),  $\Delta$ accounts for anisotropy parameters $\Delta = A(2S - 1) + BSN\textsubscript{nn}$, with $A = (\Delta E\textsubscript{FM} + \Delta E\textsubscript{AFM})/2S^2$ and  $B = (\Delta E\textsubscript{FM} + \Delta E\textsubscript{AFM})/N\textsubscript{nn}S^2$, with $\Delta E\textsubscript{FM}$ ($\Delta E\textsubscript{AFM}$) being the differences in energy between the in-plane and out-of-plane spin configurations in the ferromagnetic (antiferromagnetic) state, and $f$ is a function of the form $f(x) =  {\tanh}^\frac{1}{4}\left[ \frac{6}{N_{nn}} \log ( 1 + \gamma x ) \right]$, where $\gamma$ = 0.033. Our results are shown in the lower panel of Fig.\ \ref{Fig2}. Similarly to the investigation of $J$, we observe that the value of the Curie temperature obtained at the CA and (R)TPSS levels is about halved as compared the experimental benchmark, while it is only slightly underestimated when adopting gradient-corrected functionals, with the exception of the PW91 case. Also, we notice a good performance of both SCAN and DFT$+U$ functionals. Contrary to the case of $J$, however, we remark that hybrid functionals yield an excellent agreement with the experimental $T_C$, differing only by $\sim$2 K.  {We suggest that the reason for this traces back to the favorable compensation between the overestimated $J$ and the underestimated $E\textsubscript{MAE}$, as both these quantities enter Eq.\ (\ref{Cur}).} As DFT$+U$ and SCAN density-functionals lead to a superior description of both the Heisenberg exchange coupling and the Curie temperature, it is likely that the resulting magnetocrystalline anisotropy energies should be reliable as well. Hence, we anticipate a $E\textsubscript{MAE} \approx -0.87$ meV for ferromagnetic single-layer CrI$_3$.  

\section{Conclusion} 

In summary, we have combined  {CASSCF and MRCI calculations} with model Hamiltonians  {to quantify} the spin interactions  in two-dimensional CrI$_3$.  We have found that the inter-site magnetic interactions are primarily dictated by the ferromagnetic Heisenberg exchange coupling $J = -1.44$ meV,   {as inter-site magnetic anisotropies  $\bar{\bar{\Gamma}}$ play a practically negligible role}. Furthermore, our calculations indicate that single-layer CrI$_3$ features an out-of-plane easy axis, as confirmed by the determination of single-ion anisotropy $A = -0.10$ meV,  {$g$-tensor $g_{xx}=g_{yy}=1.90$ and $g_{zz}=1.92$}, as well as first-principles calculations of the magnetocrystalline anisotropy energy. In addition, we have assessed the performance of various flavors of popular density-functionals against our  {accurate MRCI calculations} and available experimental data,  {and found that DFT+$U$ (with $U$ = $1.0 - 1.5$ eV) and SCAN functionals shows an excellent description of exchange interactions}.  Overall, our work provides firm theoretical ground to recent experimental observations, unveils the magnitude of several magnetic interactions, and establishes reference results for future DFT studies, thereby offering a comprehensive picture of the microscopic spin physics in monolayer CrI$_3$.

\section{Acknowledgments} 

M.P.\ and R.Y.\ gratefully acknowledge Vamshi M.\ Katukuri at the Max-Planck Institute for Solid State Research (Germany) for fruitful discussions. This work was financially supported by the Swiss National Science Foundation (Grants No.\ 162612 and 172543) and the  {Sinergia Network NanoSkyrmionics (Grant No. CRSII5-71003)}. Calculations were performed at the Swiss National Supercomputing Center (CSCS) under the project s832. 

\section{Appendix A: CASSCF and MRCI Calculations } 
Many-body wavefunction calculations on finite-size model systems have been carried out with the help of {\sc molpro} quantum chemistry package \cite{Molpro12}.  For the two-site calculations, all-electron basis functions of quadruple-zeta quality were used for the Cr$^{3+}$ ions \cite{cr_basis} in the central unit while the bridging
I ligands were described with energy-consistent relativistic pseudopotentials along with quintuple-zeta quality basis sets for their valence shells \cite{iodine_basis}.  The remaining I atoms in the two-octahedra central region were treated with energy-consistent relativistic pseudopotentials along with triple-zeta quality basis sets  \cite{iodine_basis}. Cr$^{3+}$ sites belonging to the octahedra adjacent to the reference unit were described as closed-shell Co$^{3+}$ $t_{2g}^6$ ions, using all-electron triple-zeta basis functions \cite{cr_basis}, while for the I ligands belonging to these adjacent octahedra which are not shared with the central unit we relied on energy-consistent relativistic pseudopotentials along with double-zeta quality basis sets \cite{iodine_basis}. As a first step, CASSCF calculations \cite{Helgaker2000} were carried out for an average of one septet, quintet, triplet, and singlet states, essentially of $t_{2g}^3-t_{2g}^3$ character. Since CASSCF calculations also account for super-exchange processes of 
$t_{2g}^4-t_{2g}^2$ type in addition the $t_{2g}^3-t_{2g}^3$ direct exchange between nearest neighbors, the corresponding wavefunctions encodes a finite-weight contribution from inter-site excitations of $t_{2g}^4-t_{2g}^2$ type. With the CASSCF wavefunctions at hand, we next accounted for single and double excitations from the Cr $d$ ($t_{2g}$) and bridging I $p$ valence orbitals through MRCI calculations. One septet, quintet, triplet, and singlet states were considered in the spin-orbit treatment, in both CASSCF and MRCI calculations.  Finally, the resulting quantum chemistry total energies associated with one septet, one quintet, one triplet, and one singlet along with their corresponding wavefunctions were mapped onto the effective spin Hamiltonian given in Eq.\ (1) of the main text using the procedure detailed in Ref. \cite{Bogdanov13}. This involves all sixteen spin-orbit states associated
with the different possible couplings between the two nearest neighbors spins.   

In the case of the one-site calculations, all-electron basis functions of quadruple-zeta quality were used for the Cr ion \cite{cr_basis} in the reference unit while the I ligands were described using energy-consistent relativistic pseudopotentials along with quadruple-zeta quality basis sets for their valence shells \cite{iodine_basis}. The transition metal and ligand sites in the nearest-neighbor octahedral units were described analogously to the case of two-site calculations discussed above. The $g$-factors and $A$ were obtained using the wavefunctions optimized for an average of low-lying seven quartets and five doublet states at the CASSCF level.

\section{Appendix B: DFT Calculations } 
Density-functional calculations on periodic models have been carried out with the Vienna \emph{Ab Initio} Simulation Package (VASP) \cite{Kresse93, Kresse96}. Spin-orbit coupling was included in all calculations in a self-consistent manner. Electron-core interactions are described with the projector-augmented wave method, while the Kohn-Sham wavefunctions for the valence electrons were expanded in a plane wave basis with a cutoff on the kinetic energy of 500 eV. Integration over the first Brillouin zone was carried out using a mesh of 15 $\times$ 15 $k$-points for all the adopted exchange and correlation functionals but the hybrid ones, where a reduced mesh of 8 $\times$ 8 $k$-points was used. DFT+$U$ calculations were performed on top of the PBE functional, introducing an increasing amount of Coulomb on-site repulsion $U$ on the $d$ shell of Cr atoms following the rotationally invariant scheme proposed by Dudarev \cite{DFTU}. For each functional and magnetic configuration, we performed geometry optimization by relaxing the atomic coordinates until the maximum component of the Hellmann-Feynman forces was smaller than 0.005 eV/{\AA} while constraining the lattice constant to the experimental value of 6.867 {\AA}. For hybrid functional calculations, we rely on the atomic models relaxed at the GGA level. A vacuum region 17 {\AA} thick is included to avoid artificial interactions between periodic images.

 \bibliography{References}
 \end{document}